\documentclass[conference]{IEEEtran}
\newcommand{\beginbsec}[1]{\vspace{3pt}\noindent\textbf{#1. \hspace{2pt}}}
\usepackage{amsmath,amssymb,amsfonts}
\usepackage[capitalize, noabbrev, nameinlink]{cleveref}
\usepackage{graphicx}
\usepackage{textcomp}
\usepackage{xcolor}
\usepackage{url}
\usepackage[numbers]{natbib}
\def\BibTeX{{\rm B\kern-.05em{\sc i\kern-.025em b}\kern-.08em
    T\kern-.1667em\lower.7ex\hbox{E}\kern-.125emX}}

\usepackage{tikz}
\usepackage{comment}
\usepackage{amsmath}
\usepackage{algorithm}
\usepackage{algpseudocode}
\usepackage[normalem]{ulem} %for sout
\usepackage{soul}
\usepackage{pifont}
\usepackage[caption=false, font=footnotesize]{subfig}
\usepackage{tabularx}
\usepackage{multirow} 
\definecolor{OliveGreen}{rgb}{0,0.6,0}
\usepackage{arydshln}
\makeatletter
\def\hlinewd#1{%
\noalign{\ifnum0=`}\fi\hrule \@height #1 %
\futurelet\reserved@a\@xhline}
\makeatother

 \newcommand*{\RELEASE}{}
\ifdefined\RELEASE
    \newcommand\old[1]{}
    \newcommand\boris[1]{}
    \newcommand\david[1]{}
    \newcommand\dilina[1]{}
    \newcommand\shyam[1]{}
    \newcommand\rakesh[1]{}
    \newcommand\antonis[1]{}
    \newcommand\new[1]{}
    \newcommand\ant[1]{}
    \newcommand\prio[1]{}
    \newcommand\todo[1]{}
    \newcommand\invisible[1]{}

\else
    \newcommand\old[1]{}
    \newcommand\boris[1]{{\color{blue}[Boris]: #1}}
    \newcommand\david[1]{{\color{orange}[David]: #1}}
    \newcommand\dilina[1]{{\color{purple}[Dilina]: #1}}
    \newcommand\shyam[1]{{\color{gray}[Shyam]: #1}}
    \newcommand\marios[1]{{\color{green}[Marios]: #1}}
    \newcommand\rakesh[1]{{\color{magenta}[Rakesh]: #1}}
    \newcommand\prio[1]{{\color{red}[Prio!]: #1}}
    \newcommand\antonis[1]{{\color{cyan}[ak]: #1}}
    \newcommand\new[1]{{\color{red}[new]: #1}}
    \newcommand\ant[1]{{\color{cyan}[ak]: #1}}
    \newcommand\todo[1]{{\color{orange}[TODO]: #1}}
    \newcommand\invisible[1]{}
\fi

\newcommand{\coolName}{Flare}
\newcommand*\circled[1]{\tikz[baseline=(char.base)]{
            \node[shape=circle,draw,inner sep=1pt] (char) {#1};}}

\begin{document}

%%
%% The "title" command has an optional parameter,
%% allowing the author to define a "short title" to be used in page headers.
\title{Flare: Leveraging Serverless Elasticity to Absorb Microservice Load Spikes}
\author{
\IEEEauthorblockN{Dilina Dehigama\textsuperscript{1}, Shyam Jesalpura\textsuperscript{1}, David Schall\textsuperscript{2}, Antonios Katsarakis\textsuperscript{3,*},\\
Marios Kogias\textsuperscript{4}, Rakesh Kumar\textsuperscript{5}, Boris Grot\textsuperscript{1}}
\IEEEauthorblockA{\textsuperscript{1}University of Edinburgh, UK \quad
\textsuperscript{2}TU Munich, Germany \quad
\textsuperscript{3}Huawei Research, Edinburgh, UK}
\IEEEauthorblockA{\textsuperscript{4}Imperial College London, UK \quad
\textsuperscript{5}NTNU, Norway}
\IEEEauthorblockA{Email: \{dilina.dehigama, boris.grot\}@ed.ac.uk, s.jesalpura@gmail.com, david.schall@tum.de,\\
antoniskatsarakis@yahoo.com, m.kogias@imperial.ac.uk, rakesh.kumar@ntnu.no}
\IEEEauthorblockA{\textsuperscript{*}This work began when the author was at the University of Edinburgh.}
}
%% This command processes the author and affiliation and title
%% information and builds the first part of the formatted document.
\maketitle

\begin{abstract}

Online services strive to maintain application responsiveness even when the traffic is unpredictable and fluctuating. Today's online services are commonly deployed as chains of microservices, each microservice packaged as one or more containers inside virtual machines (VMs). While performant and affordable when the load is steady, VM-based deployments are known to be slow to scale when the load spikes, resulting in degraded performance for end-users of the service. To avoid such performance degradations, service providers can over-provision their deployments; however, such a strategy is costly and inefficient, leaving resources under-utilized for extended periods.

To address the challenge of unpredictable load spikes, we propose \coolName{}, a hybrid microservice architecture that combines VMs with serverless computing. \coolName{}  utilizes VMs to cost-effectively handle steady workloads and leverages serverless elasticity to absorb traffic spikes. When a spike occurs, \coolName{}  detects which specific service(s) are overloaded and shifts the excess load of only those services to serverless, thus minimizing the cost overhead.
\coolName{}  seamlessly integrates into existing auto-scaling and serverless infrastructure, requiring minimal changes to the control plane and no modifications to the application. 
% \david{Compared to an all-VM deployment, which violates an SLO target of 400ms for up to several minutes during a sudden load spike until the auto-scaler spawns new instances, \coolName is highly effective in absorbing the spike within a few seconds. It eliminates all SLO violations at the 50th percentile and even the vast majority at the 95th percentile, at a small average cost increase of less than 5\%}
% When compared to an all-VM deployment with auto-scaling, \coolName{}  achieves an average of 49.7\% reduction in peak tail latency at an average cost overhead of under 4.1\%.
When compared to an all-VM deployment with auto-scaling, \coolName{}  achieves an average of 49.7\% reduction in peak tail latency at a minor cost overhead of under 4.1\% on average.
\end{abstract}

\section{Introduction}
Today's online services are complex, tiered applications composed of multiple functionalities that must deliver a cost-effective and seamless end-user experience on a tight latency budget. 
Examples of such services include social networks, online stores, and media portals. Due to the need for high scalability, availability, and developer productivity, online services are typically developed and deployed as {\em microservices} -- a collection of lightweight independent services that communicate via remote procedure calls. 

Prior works have shown that it is common for online services to experience load fluctuation, both at regular intervals (e.g., higher load during a day and lower load at night-time) as well as less-predictable episodic spikes~\cite{alibaba-trace-analysis,madu,twitter:trace}. The latter may arise due to a major news event, an online flash sale, or the release of a suddenly popular digital media item. Regular load fluctuations are straightforward to accommodate by provisioning capacity for the expected load ahead of time. In contrast, irregular changes in load may present a challenge, especially if they are sudden and if the amplitude of the spike is large. 

Maintaining a highly responsive service in the face of a load spike is a well-known challenge \cite{firm,alibaba-trace-analysis,cbs-sports}. First, a spike must be detected and confirmed to be non-transient, and then additional resources must be provisioned and brought online before traffic can be redirected to them. In practice, these steps may take minutes or even tens of minutes, during which the service quality may be compromised. 

Indeed, Netflix reports that when its service became overloaded due to the load doubling within a span of 10 seconds, it took 5 minutes to fully restore latency Service Level Objective (SLO) through autoscaling~\cite{netflix-traffic}. In another example,  Unity \cite{unity}, a service provider that hosts several e-stores that sometimes feature flash sales, has an online forum where users frequently complain about their inability to access a given store whenever a flash sale is in progress. In one such discussion thread, titled "Store Server Overloaded Resulting in Missed Flash Sales Purchases", a user writes: "In 12 minutes since the sale started, the page has only pulled up once."~\cite{unity-forum}.

 To avoid the slow scale-out problem, services can be over-provisioned by deploying more instances than required for a given load level. However, our analysis of a week-long trace from Twitter~\cite{twitter:trace} shows that the load spikes by over 2x on several occasions during the week. Having enough stand-by capacity to absorb such spikes would be prohibitively expensive, since the extra resources would have to be deployed and paid for continuously, even when they are not needed. Another option is to predict traffic fluctuations and scale the resources proactively~\cite{madu,deepscaling}. While such predictive scaling can accurately anticipate regular load fluctuations, real-world experience shows that load may spike unexpectedly, rendering predictive scaling techniques ineffective.

In theory, the bulk of an online service can be deployed using serverless, which is known to be highly scalable both in terms of time to launch an instance (typically just seconds or less) and the number of concurrent instances (hundreds or even thousands)\footnote{Services that are inherently stateless are the ones that can be most directly ported to serverless. Fortunately, these comprise the bulk of online services deployed using microservices~\cite{porting-to-serverless}}. In practice, however, the cost of running even a moderately popular online service on serverless would be exorbitant. Our calculations show that a representative hour-long steady-load fragment of the Twitter trace would cost 2.4x more to serve using serverless than using virtual machine-based microservices.

This creates a fundamental conundrum: existing VM-based deployments are cost-effective but slow to scale, while serverless platforms offer rapid elasticity at an exorbitant continuous cost. To bridge this gap, we propose ~\coolName{}, a novel microservice architecture that dynamically combines the cost-effectiveness of VMs with the rapid scaling capabilities of serverless computing. While prior hybrid approaches exist~\cite{splitserve, cackle, mark}, they typically target specialized, batch-oriented domains like data analytics or machine learning. They lack the flexibility required for user-facing, latency-sensitive microservices that enforce strict millisecond-level SLOs. ~\coolName{} overcomes this by operating in real-time at the fine-grained service level; when a spike occurs, it identifies exactly which specific services are overloaded and selectively routes only the excess traffic to serverless functions, preserving SLOs while minimizing cost overhead.

Our evaluation demonstrates that \coolName~indeed achieves the best of both serverless and VMs -- high scalability, responsiveness {\em and} cost-efficiency. \coolName~ eliminates all SLO violations at the 50th percentile latency and significantly reduces them at the 95th percentile latency, with only a minor average cost increase of less than 4.1\%. We further show that \coolName~can be used to rapidly absorb load in case of a node failure or a spot instance being shut down, thus preserving SLO despite the loss of a node.

In short, this paper makes the following contributions: 
\begin{itemize}

    \item We analyze state-of-the-art reactive and predictive autoscaling approaches and show that neither is able to effectively deal with unexpected load spikes, corroborating industry reports (\cref{sec:motivation}). Load spikes can be mitigated by running a microservice using serverless instances but at an exorbitant cost overhead (\cref{sec:serverless}).
    
    \item We propose \coolName{}, a hybrid microservice architecture that incorporates dynamic traffic shifting from VMs to serverless functions. By selectively offloading traffic, \coolName{} improves load resilience and scalability for online microservice-based applications in the face of load spikes. \coolName{}  seamlessly integrates with existing serverless and auto-scaling infrastructure, making it readily deployable in today's clouds. (\cref{sec:design})

    \item We demonstrate that \coolName~reduces peak tail latency by 49.7\% on average and eliminates all SLO violations at 50th percentile latency while incurring an average cost increase of mere 4.1\%. \coolName~ can quickly take on additional load if a node fails, allowing it to maintain service level objectives (SLOs) despite losing that node. (\cref{sec:evaluation})%\boris{RESULTS SUMMARY; possibly comment on the node failure scenario} (\cref{sec:evaluation})

\end{itemize}

\section{Motivation}
\label{sec:motivation}

\subsection{Modern Online Services}
\label{sec:motivation:online}

Modern online services have evolved into complex systems that must be highly available, scalable, and cost-effective. Due to their user-facing nature, meeting SLOs in the form of throughput and latency constraints is crucial, as any deviation may significantly compromise the user experience~\cite{sora,SLA-cloud-autoscaling}. However, the interaction with end-users is characterized by varied and unpredictable load patterns~\cite{twitter:trace}, which makes meeting SLOs in a practical and cost-effective manner a challenge. 

To address these challenges, large service providers such as Airbnb, Netflix, LinkedIn, Uber, and Twitter have adopted a distributed microservice architecture \cite{airbnb, netflix, linkedin, uber-blog, twitter}.
In the microservices architecture, an application is composed of small, loosely coupled services, each responsible for handling specific isolated functionality. These services can be broadly categorized into stateless and stateful services, depending on their need to store and maintain state information. As a design principle, the services are designed to be stateless in nature~\cite{night-core,deathstarbench}, facilitating greater scalability. Any necessary state is decoupled from the stateless components through the use of dedicated external data stores.

\subsection{Microservice Deployment and Auto-scaling}
\label{sec:motivation:deployment}

Microservices are commonly deployed as containers on top of a cluster of VMs with the help of container orchestration platforms such as Kubernetes (K8s). \cite{k8s,dockerswarm,nomad}. K8s streamlines the management of microservices by offering features like automated deployment, scaling, and monitoring. In the K8s framework, the fundamental unit of deployment and scaling is a {\em pod}, which typically comprises a primary container and, if necessary, additional helper containers. One or more pods are deployed within a single VM on a physical node.
% \sout{These pods are deployed within VMs.}

At the initial scaling level, operations are conducted at the pod granularity, and the Horizontal Pod Autoscaler (HPA) \cite{hpa} plays a pivotal role in this process.
In the scaling-out process, the HPA orchestrates the creation of new pods within a running VM based on observed metrics, such as CPU utilization or requests per second (RPS).

In scenarios where the existing cluster of VMs lack the necessary capacity to accommodate new pods, the Cluster Autoscaler (CA)~\cite{ca} serves as the second level of auto-scaling. The CA dynamically provisions new VMs and seamlessly integrates them into the cluster, thereby enabling HPA to place additional pods inside the newly-created VMs.

\subsection{Load Variability}
\label{sec:motivation:load}

\newcommand{\spikeAmount}{2x}
The load, or request arrival rate, of a service can exhibit significant variability, influenced by factors such as the time of day, day of the week, or season \cite{twitter:trace}. This is evident in online services such as a social networking platform facing higher loads during the day and lower loads at night or a shopping website encountering increased demand during the holiday season.

\begin{figure}[t!]
    \centering
    \includegraphics[width=0.48\textwidth]{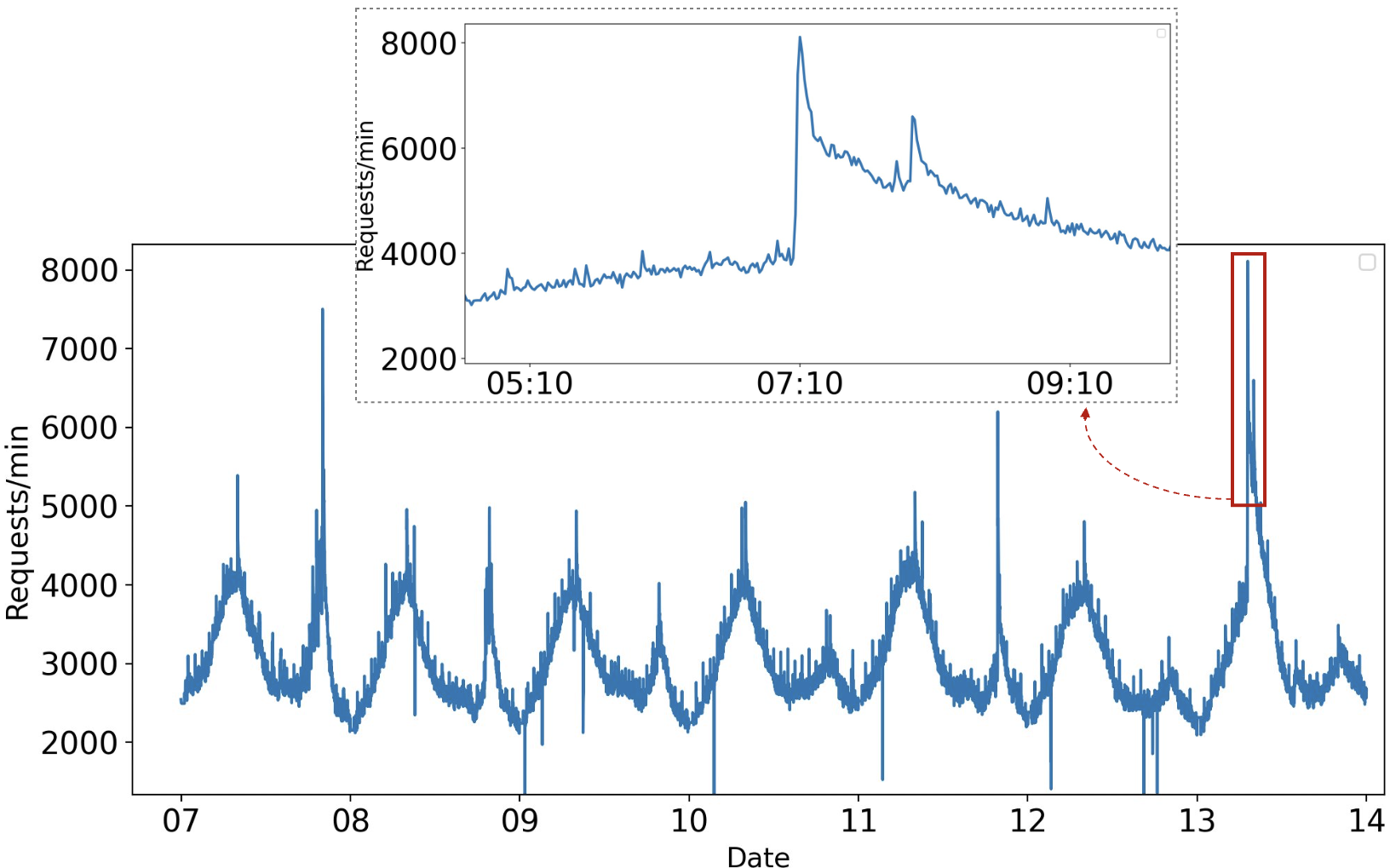}
    \caption{The load trace of Twitter over a week's time-span. The highlighted section represents a 4 hour chunk from a day with unexpected load spike.}
    % \boris{Figures should be only at the top of the page - not in the middle or bottom.}
    \label{fig:load_week}
\end{figure}

\cref{fig:load_week} illustrates the load pattern of Twitter \cite{twitter:trace} over a week-long period.
We observe a mostly consistent traffic pattern characterized by predictable periodic trends with minor fluctuations for a given time of day. However, the figure also shows multiple instances of load spiking at various points in the week without a clear periodic trend. The zoomed-in portion of \cref{fig:load_week} shows one such spike, starting from 07:10 when the load suddenly doubles from just under 4000 requests/min to 8000 requests/min. This emphasizes that microservice systems must be capable of gracefully handling unexpected spikes without compromising service quality. The question we ask in this work is: 
{\em can existing microservice deployments and their autoscaling mechanisms effectively contend with unexpected load spikes while being cost-effective?} 

% \sout{The highlighted section in \boris{Figure}~\ref{fig:load_week} zooms in on a few hours from a particular day, revealing a significant spike in demand. Notably, a similar pattern is evident even when considering longer time spans, making this a representative sample. While the majority of load variations changes gradually, at 07:00, there is a notable increase in the request arrival rate, spiking from} ..

\subsection{VM-based Microservices Meet Load Spikes}
\label{subsec:ms-meets-load-spike}

Handling unexpected load spikes poses a significant challenge for current microservice auto-scaling systems. The difficulty arises from two primary factors. Firstly, there is a \textit{detection lag}, which refers to the time it takes for the autoscaler to recognize that a genuine spike in load is occurring and is not just a temporary fluctuation. Secondly, once the spike is detected, there is a \textit{reaction lag,} which is the subsequent time needed to provision and scale out additional instances to handle the increased load.

The detection lag is caused by the autoscaler's reliance on metrics and thresholds to identify load spikes. These metrics, such as CPU utilization or request rates, are typically averaged over a specific time window. This averaging process is necessary to filter out transient fluctuations and ensure that the autoscaler reacts to sustained load increases. However, it also means that the autoscaler may not react immediately to sudden spikes in load, leading to a delay in scaling out instances.
In the K8s infrastructure, HPA's default detection lag is 15 seconds, which is the lowest HPA detection interval that can be configured including in production clusters, such as Amazon Elastic Kubernetes Service (EKS) \cite{eks-hpa-delay} and Google Kubernetes Engine (GKE) \cite{gke-hpa-delay}.

The reaction lag is the time it takes to provision new instances and deploy them within the cluster. This process can take multiple minutes, depending on the cloud provider and the size of the VMs \cite{vm-startup-time}. During this time, the system may experience performance bottlenecks as the existing instances struggle to handle the increased load while new instances are still being provisioned.  

The combined detection and reaction lag can lead to significant request queuing within existing instances, adversely affecting system performance and end-user experience~\cite{monzo}.
In essence, such reactive auto-scaling struggles to swiftly adapt to sharp spikes in workload, leading to likely SLO violations during surges in demand.

We study the impact of the load spikes on the end-to-end latency, which is defined as the total time taken from the moment a client's request is sent to when the response is received. The experiment employs the BookInfo microservice-based application from Istio \cite{istio-home}, comprising five microservices implemented in various language runtimes \cite{bookinfo}. The traffic is generated by replaying an hour-long load trace containing the highlighted load spike in \cref{fig:load_week}. The application is deployed on K8s in an AWS EKS cluster that integrates both HPA and CA, representing state-of-the-art production-grade auto-scaling. \cref{sec:methodology} provides detailed information on the parameters used in this study.

\begin{figure}
    \centering
    \includegraphics[width=0.49\textwidth]{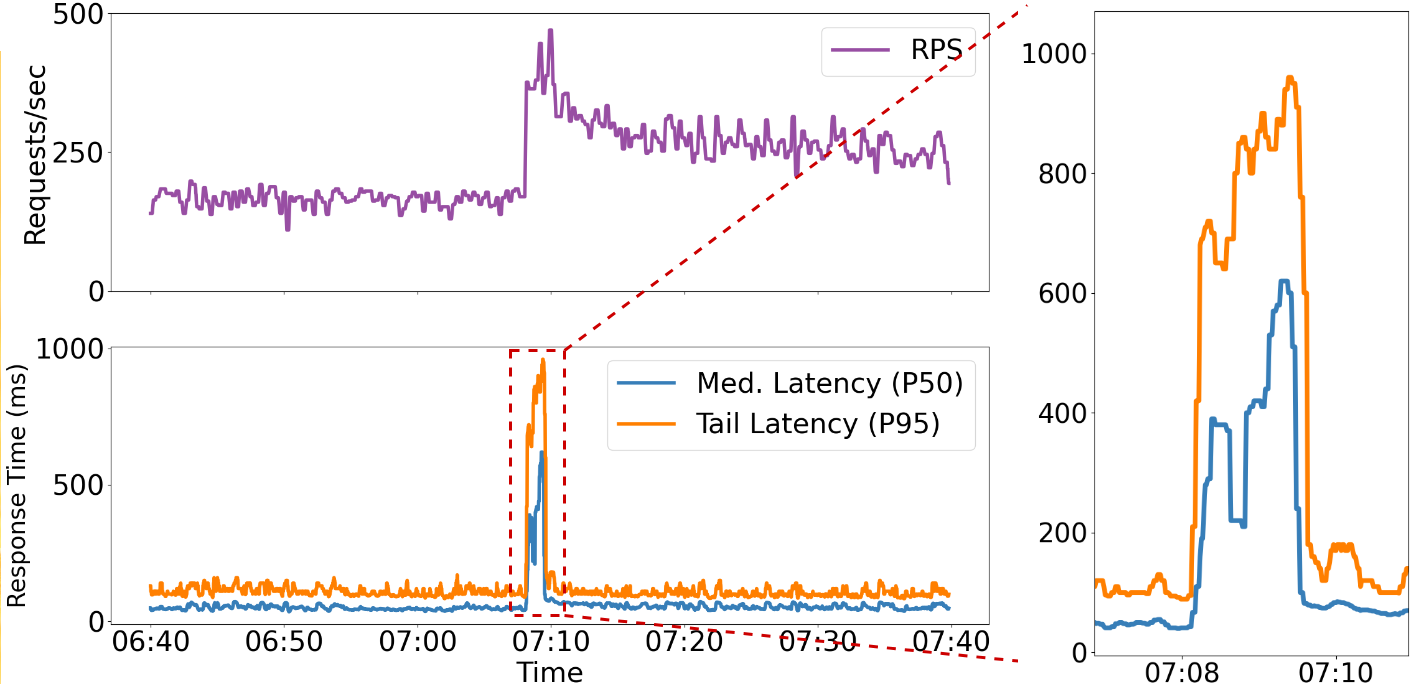}
    \caption{Impact of a sudden spike in load on a VM based microservice application.}
    \label{fig:microservices+spike}
\end{figure}

\cref{fig:microservices+spike} shows the load and resulting end-to-end latency over time.
The top graph reveals an initially stable request rate of around 200 RPS, followed by a load spike that causes the RPS to exceed 400 at around 07:08, 
% \rakesh{Based on the zoomed out portion of the figure, it should be at 07:08.}
after which the RPS gradually declines. The bottom graph shows the corresponding median (P50) and tail latency (P95). Both latencies are low and stable until the traffic spikes at around 07:08, at which point latencies surge until the median latency peaks at over 600ms and the tail latency exceeds 900ms, representing an increase of 8.2x \& 9.6x, respectively, compared to pre-spike latencies. 

Latencies return to pre-spike levels just after 07:10, indicating that the combined detection and reaction time of the system is around two minutes. These findings are in line with industry reports; for instance, Netflix notes that when its service became overloaded due to the load doubling within a span of 10 seconds, it took 5 minutes to fully restore latency through auto-scaling~\cite{netflix-traffic}. 

In order to avoid the problem of slow auto-scaling under a sudden load surge, service providers often over-provision their clusters \cite{burscale}.
However, over-provisioning can be a prohibitively expensive strategy if load fluctuates unexpectedly and by a large amount.
For instance, given that the magnitude of the spike illustrated in \cref{fig:load_week} is twice that of the steady-state load beforehand, to fully absorb the spike and avoid any latency impact, the resources would need to be over-provisioned by a factor of two, correspondingly increasing the cost.
Clearly such an over-provisioning strategy is extremely wasteful in terms of resources (which would need to be idle most of the time) and money.

\subsection{Proactive Resource Provisioning}
\label{subsec:proactive}

    \begin{figure}[!t]
    \subfloat[A day with unexpected load spike \label{fig:mispredicted1}] {
        \includegraphics[width=0.49\textwidth]{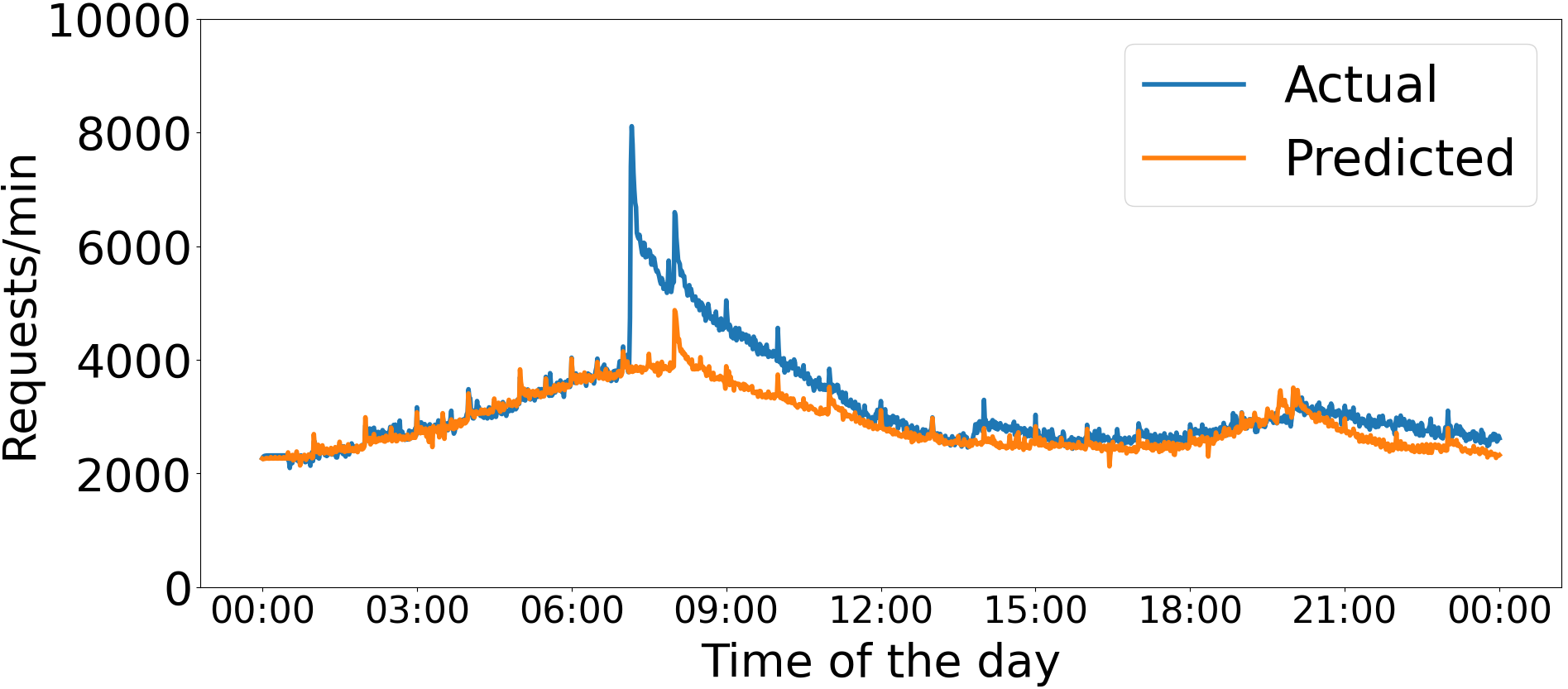}}
        
    \centering
    \subfloat[Another day with unexpected load spike \label{fig:mispredicted2}] {
        \includegraphics[width=0.49\textwidth]{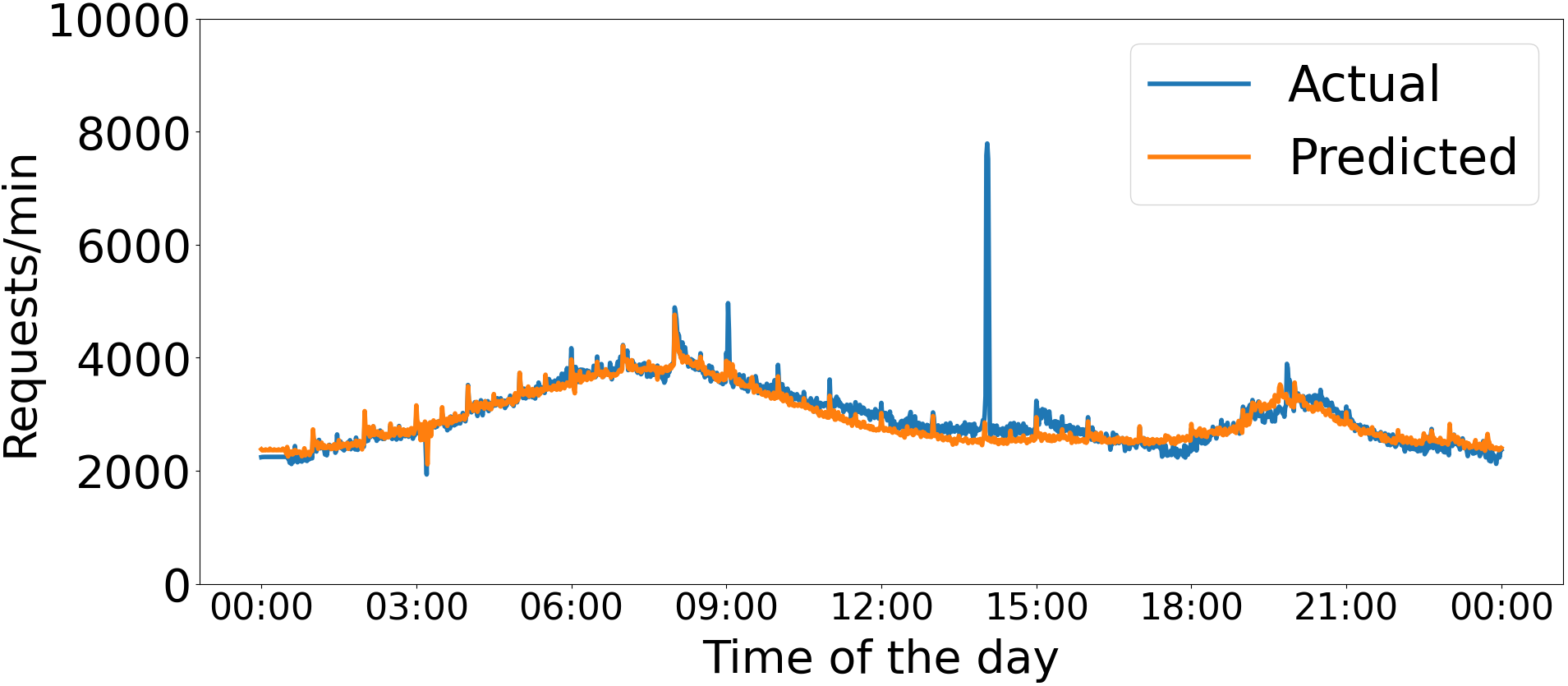}}
       
    \caption{Load prediction for two days with unexpected load spikes}
    \label{fig:mispredicted}
\end{figure}

Instead of reacting to sudden load spikes, prior works have proposed proactive auto-scaling mechanisms that aim to predict future load patterns and provision resources accordingly to preemptively handle workload fluctuations\cite{carvalho, self-driving, optimus, Autopilot, cloudscale}. These approaches typically utilize machine learning models to analyze historical data and forecast future resource requirements \cite{deepscaling,madu}. However, despite generally good accuracy, proactive auto-scaling approaches are challenged by sudden and unexpected load spikes. 

To evaluate the effectiveness of proactive models, we employed a widely-used machine learning model called Seq2seq \cite{seq2seq}, which is utilized in the state-of-the-art proactive auto-scaling mechanism at Alibaba \cite{madu}. We train the Seq2seq model using three months of Twitter trace data to evaluate its prediction accuracy.

Figure \ref{fig:mispredicted} presents two day-long periods featuring unexpected load spikes, comparing the load predicted by Seq2seq (orange line) versus the actual load (blue line). We observe that during portions with regular load (specifically before 07:00 and after 12:00 in Figure~\ref{fig:mispredicted1}, and outside the isolated surge around 14:00 in Figure~\ref{fig:mispredicted2}), Seq2seq achieves good prediction accuracy, maintaining less than a 14\% deviation. Conversely, during the spike periods, the predicted load significantly deviates from the actual load, with maximum deviations exceeding 200\% and 300\% in Figures~\ref{fig:mispredicted1} and \ref{fig:mispredicted2}, respectively. This highlights the limitations of predictive models in accurately forecasting unexpected load spikes, thereby hampering the effectiveness of proactive auto-scaling mechanisms in such scenarios.
\section{Can Serverless Help?}
\label{sec:serverless}

\subsection{Serverless Basics}
\label{sec:serverless:basics}

Serverless computing, or Functions-as-a-Service (FaaS), abstracts server management by executing applications as small, event-driven, stateless functions. Cloud providers dynamically scale these compute resources based on instantaneous load, eliminating manual infrastructure provisioning~\cite{serverless-computing-book}. 

While traditional VM-based microservice deployments struggle to provision resources quickly enough to handle sudden spikes in load, serverless computing offers a promising alternative. Serverless functions scale near-instantaneously to absorb these rapid traffic surges, going from zero to thousands of active instances \cite{lambda_scaling} within seconds due to lightweight and stateless nature. In addition to rapid scaling, serverless computing features a pay-as-you-go model that eliminates costs for idle resources, making it a compelling option for applications with highly variable load patterns.

\subsection{Microservices on Serverless}
\label{sec:serverless:state}

To fully exploit the benefits of serverless for an existing microservice application, each service must be migrated to serverless. For stateless services, the migration is simple as they naturally fit the serverless model. For stateful services, such as databases and caches, the migration can be more involved due to additional state transfer requirements. In practice, however, microservice developers tend to keep their individual services stateless and leverage managed database and cache services offered by cloud providers for maintaining the state due to the latter's superior fault-tolerance and scalability properties. Examples of such services include the AWS RDS (a fully managed cloud database service), DynamoDB (a fully managed NoSQL database), and ElastiCache (a fully managed object cache). The use of managed services for maintaining application state naturally facilitates deployment of microservices as serverless flows~\cite{night-core}.

Indeed, adhering to stateless design principles is a well-known best practice when designing microservices~\cite{deathstarbench}.
A recent study demonstrated the feasibility of migrating mid-tier services, which compromise the bulk of complex microservice-based applications, to serverless architectures. In these applications, over 90\% of mid-tier services implementing business logic are either inherently stateless or can be seamlessly ported to serverless environments \cite{porting-to-serverless}.

\subsection{Evaluating Scalability of Serverless}
\label{sec:serverless:scalability}

We study the effect of the load spike highlighted in \cref{fig:load_week} on the BookInfo benchmark described in~\cref{subsec:ms-meets-load-spike}, which for this study we deploy fully on serverless.  

\begin{figure}
    \centering
    \includegraphics[width=0.49\textwidth]{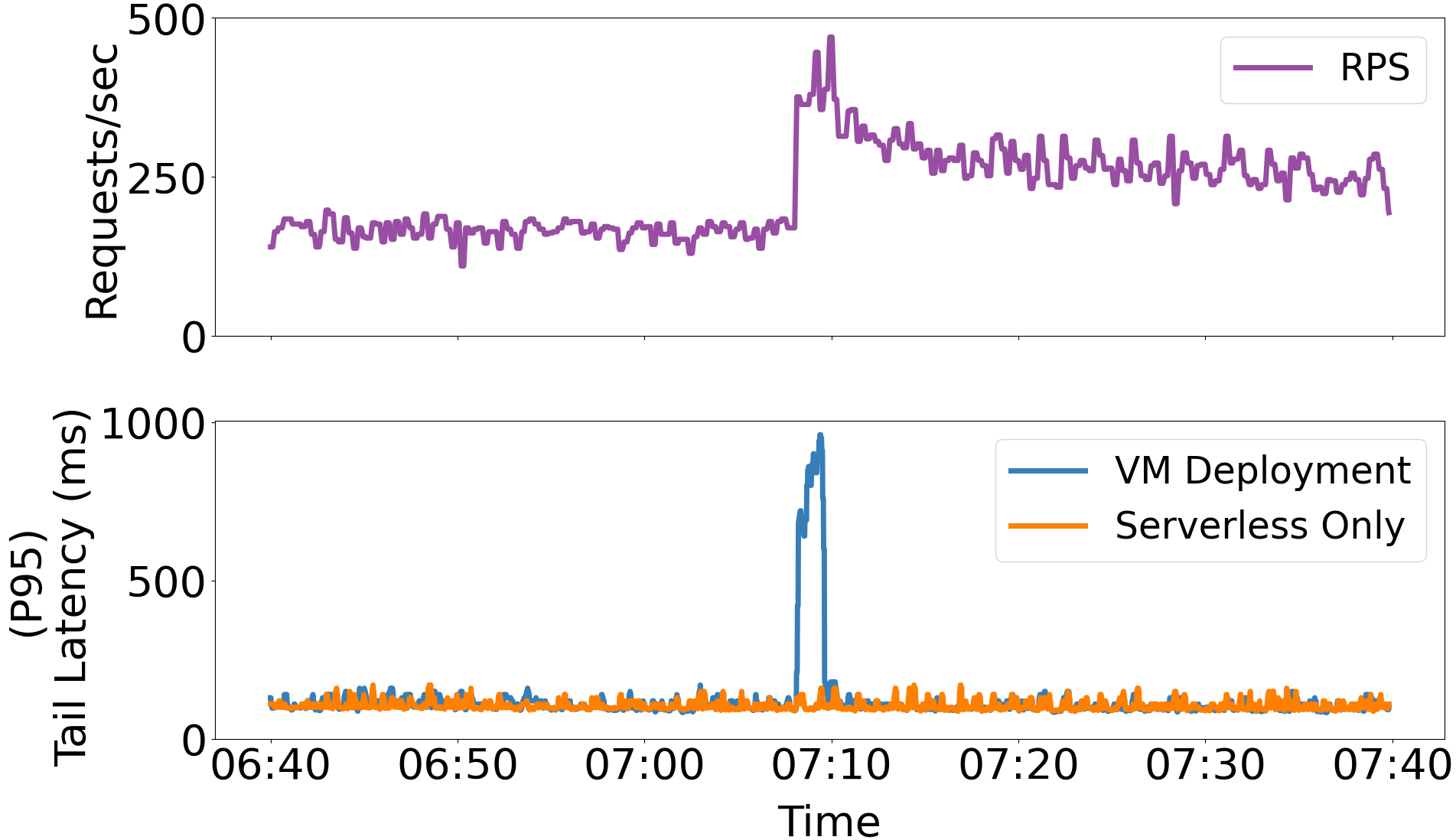}
    \caption{Impact of a sudden load spike on tail latency (P95) on a entirely serverless-function based deployment compared to VM-based deployment}
    \label{fig:serverless+spike}
\end{figure}

\cref{fig:serverless+spike} shows the results of the study. Unlike the VM-based deployment depicted in \cref{fig:microservices+spike}, where the load spike resulted in a significant increase in tail latency of 9.6x compared to pre-spike stable latency, the serverless-native deployment has absorbed the spike with virtually no impact on tail latency. This result highlights the resilience of serverless functions in reacting to unpredictable spikes in load.

Alas, the high scalability offered by serverless comes at a cost. Because serverless functions are billed on a per-execution basis, they are not cost-effective in handling a steady load compared to pre-provisioned resources like VMs~\cite{sebs,serverless-berkeley-view}. 

% To investigate the cost implications of running a microservice-based application on VMs versus on serverless, we use the BookInfo application introduced in \cref{subsec:ms-meets-load-spike}  and run in AWS on EC2 instances and on Lambdas under steady load using  an hour-long segment of the Twitter trace (from 05:40 to 06:40 in Figure~\cref{fig:load_week}).

To investigate the cost implications of running a microservice-based application on VMs versus on serverless, we conducted a cost estimation using the BookInfo application introduced in \cref{subsec:ms-meets-load-spike}. We used an hour-long segment of the Twitter trace exhibiting only modest load fluctuations (from 05:40 to 06:40 in \cref{fig:load_week}).
For a fair comparison, we did our estimation ensuring that the number of vCPUs (virtual CPUs) allocated to Lambdas matches that of EC2 instances. Note that in AWS Lambda, users cannot directly specify the number of vCPUs per function instance. Instead, they configure the amount of memory allocated to a serverless function, and based on that, AWS allocates a certain number of vCPUs~\cite{lambda-memory-cpu}. Thus, we provisioned the memory of the Lambda functions such that the total number of vCPUs allocated to those functions matches the total number of vCPUs of the VM instances. Our results show that running the hour-long relatively steady load on the target microservice-based application is 2.4x more expensive on serverless deployments (\$0.48) compared to VMs (\$0.20). Meanwhile, the difference in tail latencies (P95) between the two deployment types is minimal, at 106 ms for serverless versus 113 ms for VMs.

\subsection{Summary}
\label{sec:serverless:summary}

Complex online services typically utilize a microservice architecture for enhanced productivity, robustness, and flexible resource scaling. Problematically, VM-based microservice deployments struggle to cope with sudden load spikes, as the delay in detecting a spike and provisioning new VM instances can take minutes, leading to queued requests and increased latency. Deploying microservices using serverless technology is an effective way to mitigate unexpected load spikes but comes at a high cost overhead during periods of steady or predictable load. What is a needed is a way to combine the strengths of the two architectures -- the cost-effectiveness of VMs for steady loads and the rapid scaling capabilities of serverless for unexpected load spikes.

\section{\coolName{} }
\label{sec:design}

\subsection{Overview}
\label{sec:design:overview}

% design goals
\coolName{} is a system that achieves the best of both worlds in scalable infrastructure,
i.e. the cost-efficiency of VM deployments and the responsiveness of serverless computing.
We design \coolName{} based on the following goals.
\coolName{}:
i) can seamlessly and opaquely use both VM and serverless resources according to the system state, without exposing its internals to the user;
ii) is compatible with existing cloud infrastructure, builds on top of existing systems, and introduces minimal changes;
iii) targets generic microservices chains and is fine-grained at a microservice granularity;
iv) optimises both for cost and SLO violations.

% key idea
At the heart of \coolName{} is a simple idea of temporarily deploying serverless resources to absorb the excess traffic during load spikes
and using those resources only for the time it takes for new VMs to come online and take over.
This way \coolName{} efficiently absorbs the extra load, thus avoid prolonged periods of violated SLOs, without the prohibitive cost of always running on serverless or keeping sufficient idle VM capacity.

An important aspect of \coolName{} is that it functions at the finest granularity along the microservice chain, identifying individual service(s) that may be overloaded and redirecting individual requests that are targeting these services to serverless instances. Moreover, from a client's and application's perspective, \coolName{} is completely opaque, meaning that neither the client nor application is aware of whether a given request executed on a VM or serverless instance.

To achieve that and remain complementary to existing infrastructure, \coolName{} consists of three main components:
i) the monitoring infrastructure, ii) the load balancing infrastructure, and iii) the control plane.
For the first two components, \coolName{} leverages the existing infrastructure of the deployment environment, e.g. K8s.
The control plane introduces a novel component, the {\em \coolName{} controller}, that consumes data of the current system load from the monitoring subsystem and, if necessary, configures the load balancing infrastructure to seamlessly steer excess load to serverless resources until additional VM capacity comes online.
\subsection{System in Action}
\label{subsec:dynamic_load_shifting}

\begin{figure}
    \centering
    \includegraphics[width=0.4\textwidth]{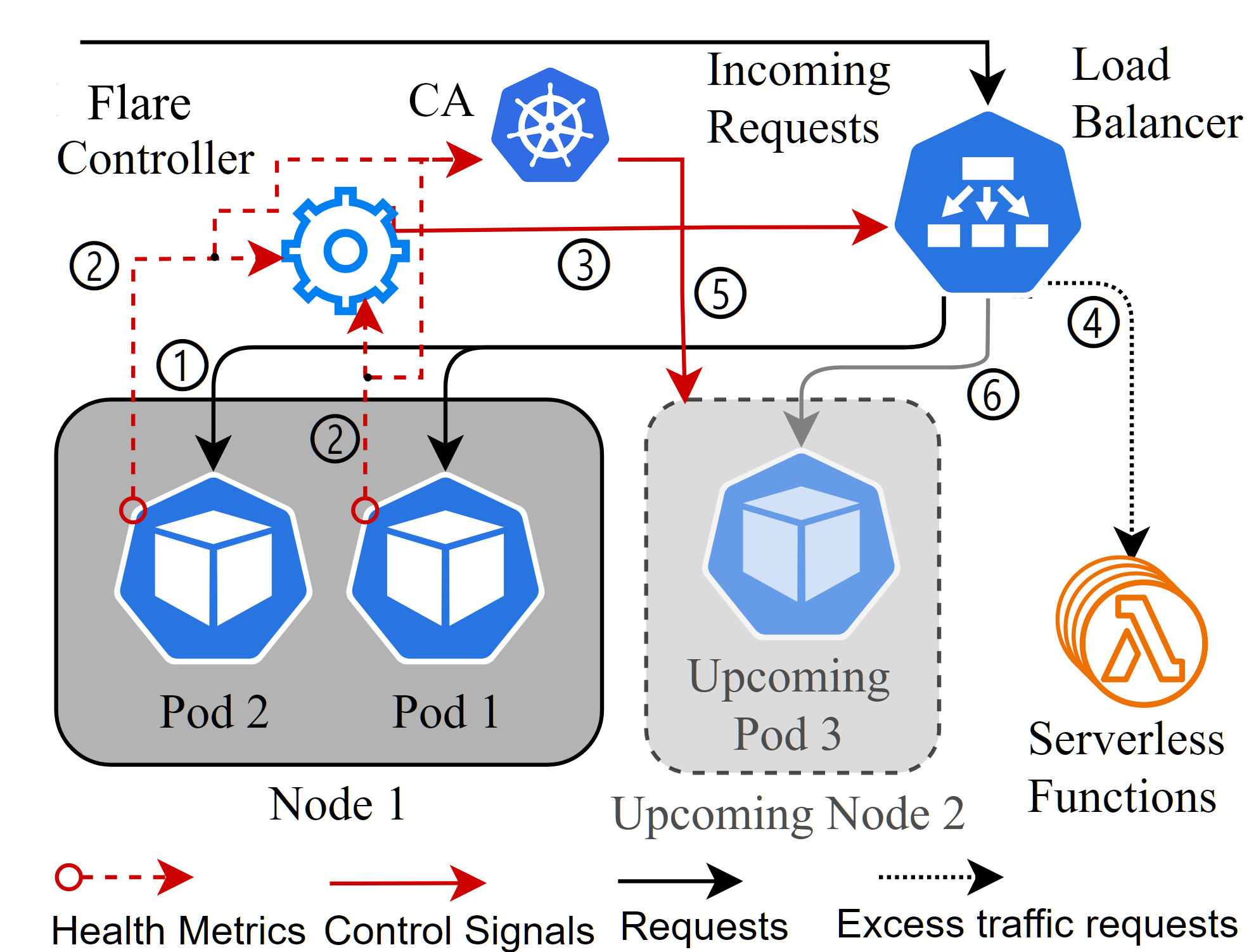}
    \caption{\coolName~design overview.}
    \label{fig:design}
\end{figure}

\cref{fig:design} focuses on the case of a single scalable microservice and provides a high-level depiction of \coolName{} in action. 
It assumes \coolName{} is deployed on top of K8s, hence it reuses its existing monitoring infrastructure.
Microservices run in pods, while an external load balancer, e.g. Envoy or AWS's Application Load Balancer ~\cite{aws-alb, envoy} steers incoming traffic.
The \textit{\coolName~Controller} runs as a microservice within the cluster and continuously monitors the RPS and CPU metrics in order to detect load spikes \circled{2}.

Under steady load, the load balancer directs 100\% of the incoming service requests to the existing pods in VMs \circled{1}.
Upon identifying a load spike, the \coolName~controller instructs the load balancer to change its load distribution and shift a fraction of the traffic to serverless functions \circled{3}.
Consequently, the load balancer routes the configured traffic portion to serverless functions \circled{4}. With excess load shifted to serverless, the load on VMs remains acceptable thus preserving SLOs.
Simultaneously, \coolName~controller does not interfere with the operation of the existing autoscaling mechanisms, e.g. Cluster Autoscaler (CA), which brings up new VMs to accommodate the increased load \circled{5}.
Once the new VMs are operational, the \textit{\coolName~controller} instructs the load balancer to shift the traffic back to the VMs \circled{6}.

\subsection{\coolName{}  Controller}
\label{subsec:hydra_controller}

The \textit{\coolName~Controller} is the brain behind \coolName{} and is in charge of two functionalities.
First, it monitors CPU and RPS metrics for all services deployed in the cluster.
Second, based on observed metrics, it configures the load balancing infrastructure to steer traffic only to VMs or to both VMs and serverless resources. Crucially, the monitoring and load-balancing happens for each individual service along the microservice chain.

\coolName{} employs a weighted load balancing approach~\cite{envoy-weighted-lb,nginx-weighted-lb} to effectively distribute traffic between VMs and serverless components. Under steady load conditions, traffic is directed solely towards VM clusters, with serverless clusters receiving no traffic. In the event of a load spike, \coolName{} dynamically calculates weights in real-time for both VM and serverless clusters based on observed metrics. During weight calculation, \coolName~ ensures that the weight assigned to the serverless cluster corresponds only to the excess traffic generated by the load spike. Once the underlying autoscaling mechanism (e.g. CA+HPA) has added new VMs to the cluster and they are operational, the \coolName~controller updates the load balancer weights to redirect traffic back to the VMs.

\section{\coolName{} Under the Hood}
\label{sec:implementation}

\subsection{Technologies Used}

\beginbsec{Service provisioning}
\coolName{} is implemented on top of Kubernetes. For serverless deployments, we use Knative, an open-source project that extends K8s to provide a framework for deploying and managing serverless functions~\cite{knative}.

\beginbsec{Service Mesh \& Virtual Services}
We employ the Istio service mesh~\cite{istio,istioclients,linkerd} with Envoy proxy sidecars~\cite{envoy,bloc,appmesh} to manage communication, observability, and routing without cluttering the underlying application code. Within this mesh, we leverage Istio Virtual Services to define routing rules, enabling dynamic traffic shifting between VMs and serverless instances.

\beginbsec{Fine-Grained Monitoring}
For real-time load spike detection, we extend the K8s CAdvisor~\cite{cadvisor} source to collect pod CPU usage statistics every 1 second. These statistics, along with service-level metrics like RPS, are consumed by Prometheus~\cite{prometheus} and exposed to the controller to enable weight calculation within the load-balancing algorithm.

\subsection{\coolName~ Controller}

\beginbsec{Integration with the Autoscaler}
In order to decide if the provisioned VM capacity is sufficient, the K8s autoscaler monitors the CPU utilization of each service and provisions more capacity if the configured threshold is exceeded. With \coolName's rapid detection and reaction, a sudden increase in load will immediately result in serverless instances getting engaged to maintain an acceptable load on the existing VMs. As a result, the autoscaler will observe a steady load on the VMs, thus preventing it from spinning up new VM instances. 

To address this issue, \coolName{} tracks the total RPS on each service across both VM and serverless instances. It then maps RPS to CPU utilization, which yields an estimate of total CPU utilization for the given service across its VM and serverless instances. Feeding this value to the autoscaler enables it to make correct decisions about scaling VM instances based on total load, even when the serverless instances are handling a portion of the traffic.

\beginbsec{Weight Calculation}
\label{sec:implementation:weight}
Weighted load balancing in \coolName~uses two weights, $W_v$ and $W_s$, denoting the proportion of traffic directed to VMs and serverless instances, respectively. At timestamp $t$, $W_v$ is calculated as the ratio of the \textit{service throughput capacity} (the maximum collective RPS all active pods can handle simultaneously) to the estimated load at $t+1$. Defining the recent change in load as $\Delta RPS = RPS_t - RPS_{t-1}$, the load for the next time step is estimated as $RPS_t + \Delta RPS$ if the load is increasing ($\Delta RPS > 0$). If the load is steady or decreasing ($\Delta RPS \leq 0$), the estimate remains $RPS_t$. Defined formally as:

$$W_v = \frac{\textit{Service Throughput Capacity}}{\begin{cases} RPS_t + \Delta RPS & \text{if } \Delta RPS > 0 \\ RPS_t & \text{if } \Delta RPS \leq 0 \end{cases}}, \quad W_s = 1 - W_v$$

\subsection{Handling Traffic Shifts in Chained Microservices}

The microservice design philosophy calls for breaking up complex functionalities into smaller services that call each other as needed, effectively forming a call chain. A complex application may have multiple such non-overlapping microservice chains. For instance, consider a trivial social media application implemented with microservices. It may deploy one microservice chain View for viewing a post, and a completely separate chain Post for posting new content. A particularly popular post may result in a flurry of views, potentially overloading one or more services along the View chain of microservices, while leaving the Post chain lightly utilized. 

A blunt way to deal with a load spike is simply to redirect all or some of the traffic of the entire application to serverless. However, that would mean also redirecting requests that do not touch the overloaded service(s), thereby incurring costs for using serverless instances unnecessarily. Instead, as explained earlier, \coolName{} identifies the overloaded service(s) and only redirects some of the traffic targeting those specific service(s) to serverless instances. The question, however, is what happens to requests that enter the serverless path -- do they return back to VM instances after the bottle-necked service is handled on serverless, or do they continue as serverless?

Clearly, redirecting requests back to VMs is the most cost-efficient solution. It is, however, also one that incurs the highest design and implementation complexity. The intricate call graph of complex applications and the dynamic nature of request paths means that, in practice, a given service may invoke one of several downstream services based on request specifics. Thus, a particular serverless instance would need to maintain load information about all of the downstream services. Alternatively, a dedicated load balancer stage would need to be added after each service, further complicating the design and increasing end-to-end latency. 

To avoid the complexity of having requests served by serverless returning back to VMs, our current implementation chooses to maintain a request within the serverless execution environment once it enters the serverless path at any point of the microservice call chain. While this design choice may incur additional costs due to the utilization of serverless instances for non-overloaded microservices further in the chain, it promotes a simplified architecture. Our evaluation of \coolName{} indicates that this design choice incurs negligible cost. This decision is also not fundamental to \coolName{} and can be revisited in a future implementation.

\subsection{Service Portability and Adaptation Effort}

Moving workloads between VMs and serverless environments is simplified as containers have become the standard unit of deployment. Most modern serverless platforms accept container images directly, enabling the exact same image to be deployed to both the VM-based Kubernetes cluster and the serverless functions. 
Although the deployment artifact remains consistent, the underlying communication protocols can vary. For example, Knative uses gRPC and HTTP while AWS Lambda has its own specific invocation model \cite{aws-lambda-event}. Such invocation discrepancies can now be easily resolved using lightweight adapters, such as the AWS Lambda Web Adapter \cite{aws-lambda-web-adapter, serverless-adapter, grpc-gateway}. The adapters function as a transparent bridge and allow standard web applications to execute seamlessly within the serverless environment. Consequently, the application logic remains untouched. The continued evolution of adapter tools facilitates greater portability across platforms and reduces the developer burden effectively.

\section{Experimental Methodology}
\label{sec:methodology}

\newcommand{\callie}{Callee}
\newcommand{\caller}{Caller}

\beginbsec{Load traces}
We use two load traces from Twitter \cite{twitter:trace} to generate the load. From traces shown in \cref{fig:mispredicted1} \& \cref{fig:mispredicted2}, we extract one hour-long portions named Trace A and Trace B that include the load spike. Since, the original trace is sampled, we scale up the trace for our evaluation to generate sufficient load to trigger CA-level auto-scaling \cite{trace-upscale}.  We use locust \cite{locust} to generate the load. Locust is deployed in a VM with 4 vCPU cores and 16 GB of RAM in the same AWS region as the cluster.

\beginbsec{Benchmarks} 
We conduct our evaluation using three microservices benchmark applications: BookInfo, Online Boutique, and Hotel Reservation \cite{bookinfo,online-boutique,deathstarbench}. These applications showcase characteristics of today's microservice deployments. They encompass multiple service chains, diverse communication protocols and are implemented in various widely-used programming languages. Thus, they serve as representative examples for our evaluation. We note that we ported one service of the BookInfo microservice application from Java to Go to mitigate Java's high cold-start latency on serverless \cite{vhive}.

For each configuration (described below), we replay the load traces on each application three times and report the results of with the median latency.

\beginbsec{Configurations}
We compare \coolName~against the K8s default baseline (\textit{Baseline}), that utilizes an EKS cluster with default parameters for HPA and CA. The HPA is configured to trigger scaling when the average CPU utilization reaches a threshold of 50\% \cite{madu}. Additionally, the average RPS threshold is configured to correspond to the RPS at 50\% CPU utilization. The HPA has a default metrics polling interval of 15 seconds (\textit{\small--horizontal-pod-autoscaler-sync-period}), while the CA checks the need for new VMs at intervals of 10 seconds(\textit{\small{-scan-interval}}). 

\beginbsec{Cluster}
The K8s cluster is deployed using Amazon EKS \cite{eks} and the nodes are EC2 instances. The nodes are of type \textit{t3.xlarge} with 4 vCPU cores and 16 GB of RAM. The cost of each VM is 0.1670 USD per hour.\cite{aws-vm-cost}. For Knative deployments, dedicated VMs are allocated within the same K8s cluster. Since the cold start times of the production serverless offerings like AWS Lambda or Azure Functions is 10 times lower than that of Knative \cite{xanadu}, we keep Knative functions warm to achieve a similar level of performance to production serverless offerings. We separately evaluate the effect of warm and cold serverless starts on \coolName's performance using AWS Lambda in~\cref{sec:evaluation:lambda}.

\begin{figure*}[!t]
    \subfloat[Zoomed Trace A for all three applications] {
        \includegraphics[width=0.48\textwidth]{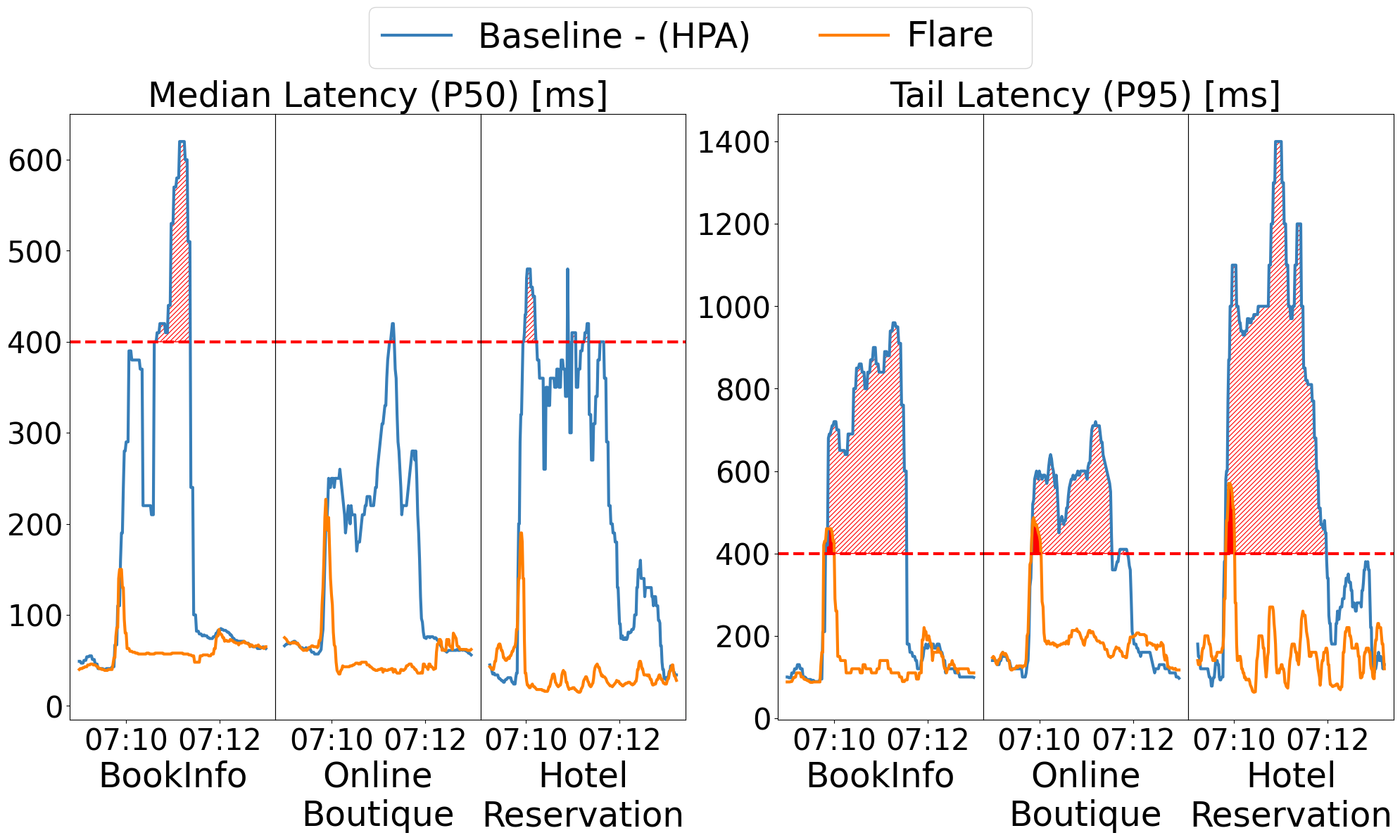}
        \label{fig:eval-all-a}}
    \hfill
    \subfloat[Zoomed Trace B for all three applications] {
        \includegraphics[width=0.48\textwidth]{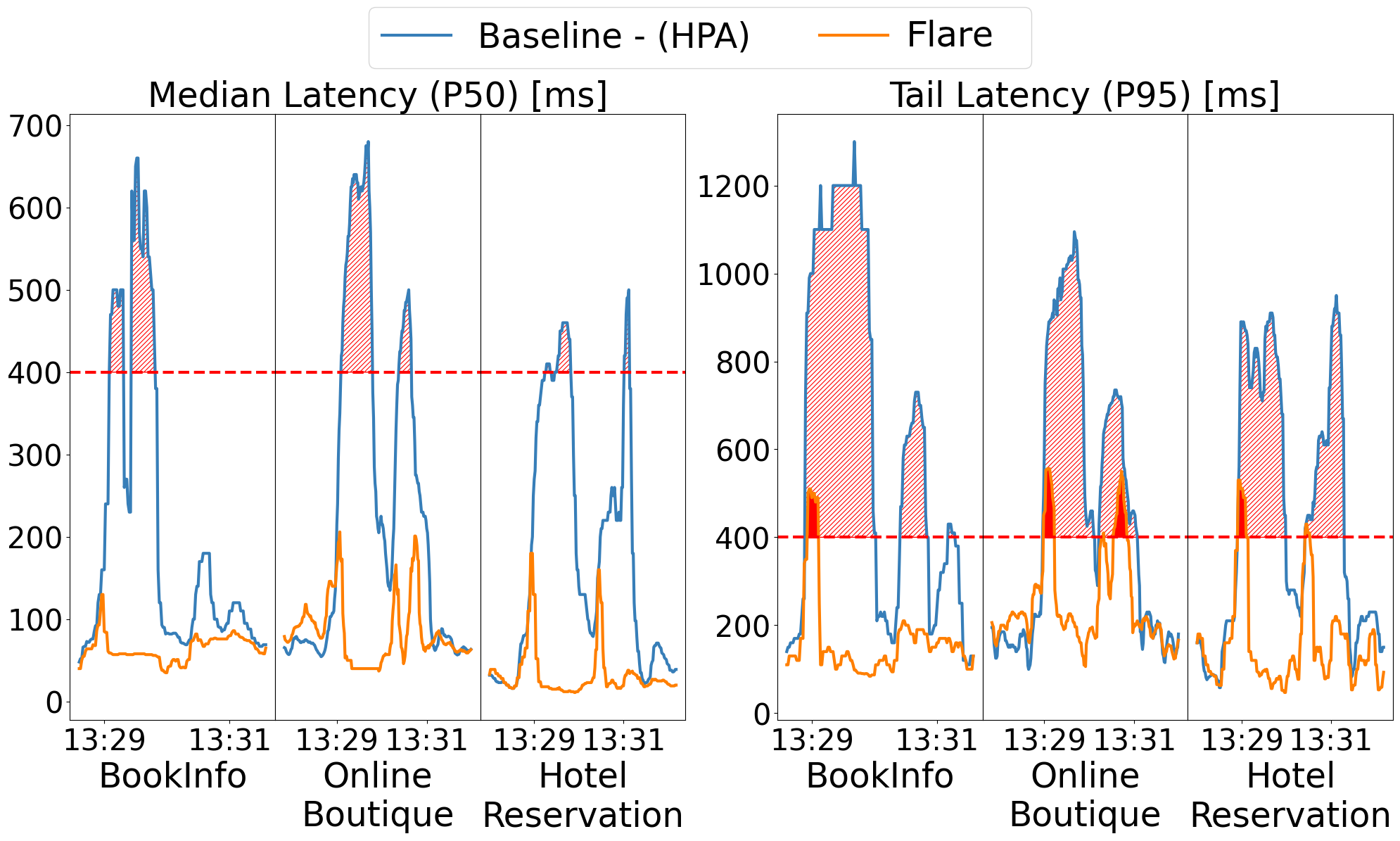}
        \label{fig:eval-all-b}}
    \caption{Latency comparison. Red dashed line marks the 400ms SLO.}
    \label{fig:eval-all}
\end{figure*}

\beginbsec{Cost Estimate}
For the serverless cost estimation, we utilized the execution time and allocated vCPU of our Knative functions. However, since AWS Lambda ties vCPU allocation directly to provisioned memory and does not allow direct CPU selection, we first determined the specific Lambda memory configuration required to yield the equivalent vCPU power of the Knative function~\cite{lambda-memory-cpu}. We then used this configuration in the AWS Lambda cost calculator~\cite{lambda-pricing-calculator}.
\section{Evaluation}
\label{sec:evaluation}
%-------------------------------------------------------------------------------

% \begin{figure*}[!t]
%     \subfloat[Zoomed Trace A for all three applications] {
%         \includegraphics[width=0.48\textwidth]{figures/perf_13_flare.png}
%         \label{fig:eval-all-a}}
%     \hfill
%     \subfloat[Zoomed Trace B for all three applications] {
%         \includegraphics[width=0.48\textwidth]{figures/perf_2_flare.png}
%         \label{fig:eval-all-b}}
%     \caption{Latency comparison. Red dashed line marks the 400ms SLO.}
%     \label{fig:eval-all}
% \end{figure*}

With \coolName{}'s evaluation we aim to answer the following questions:
\begin{itemize}
    \item What is \coolName{}'s impact on microservices' performance and SLO violations? (Section~\ref{sec:evaluation:perf})  
    \item How cost effective is \coolName{} compared to baseline? (Section~\ref{sec:evaluation:cost})
    \item How easy is it to integrate \coolName{} with commercial serverless offerings and how does it perform? (Section~\ref{sec:evaluation:lambda})
    \item Can \coolName{} also be leveraged to provide resilience against sudden node failures? (Section~\ref{sec:evaluation:failures})
\end{itemize}

\subsection{Performance}
\label{sec:evaluation:perf}

We first evaluate \coolName's ability to mitigate latency impact due to sudden load spikes. We compare the median (50\%) and tail (95\%) latencies of \coolName~ against the VM-only baseline. Evaluation is conducted on the three microservice applications and the two trace segments described in Section \ref{sec:methodology}. 
To put the measured latencies in perspective, we use 400ms as the SLO target based on Google's study that identified this value as an upper bound for a good end-user experience. Since \coolName~ engages only during the spike, the baseline and \coolName~ exhibit similar latencies before and after the spike periods. Therefore, our studies focus only on the intervals corresponding to the spike, namely 7:10-7:12 for Trace A and 13:28-13:30 for Trace B.
% The results are reported in the following order: BookInfo, Online Shop, and Hotel Reservation.

\cref{fig:eval-all-a} \& \cref{fig:eval-all-b} illustrates for all three benchmarks the end-to-end latency impact on both tail and median for Trace A and Trace B, respectively. 
We observe that in the baseline deployment, the load spikes severely impact the median and tail latencies, resulting in significant SLO violations. In Trace A, the median latency increases by 5.9-19.2x after the load spike compared to pre-spike latency, exceeding the 400ms SLO by 5-55\% for all three applications. Similarly, for Trace B, the median latency jumps up by 8.7-20x, violating the SLO for a total of 25-59 seconds.

The tail latencies (right graphs in \cref{fig:eval-all-a} \& \cref{fig:eval-all-b} follow a similar trend, with a 4.2-12.9x and 8.7-12.3x increase of the tail latency over the pre-spike level for Trace A and Trace B, respectively. 
For 5\% of the incoming requests (i.e., user queries), the acceptable limit of 400ms is exceeded by 80\% to more than 200\% for Trace A, and by 173-225\% for Trace B.

With \coolName, the excess load resulting from the traffic spike is rapidly shifted to serverless instances, drastically reducing the end-to-end latency compared to the baseline. The median latency with \coolName{} increases by 1.6-7.6x (Trace A) and 1.7-7.2x (Trace B), never violating the SLO. Similarly, the tail latency increases by 2.8-5.3x (Trace A) and 3.2-4.9x (Trace B) for a short period over the pre-spike tail latency. Compared to the VM-only baseline, this corresponds to a reduction of 32-59\% and 44-61\% in peak tail latency for Trace A and Trace B, respectively. 
Across all benchmarks, \coolName{} reduces the median latency by 66\% and the tail latency by 49.7\% compared to the baseline.
% exceeding 400ms for a short period by 15-42\% and 27-33\% for Trace A and Trace B, respectively.

% However, while \coolName{} does not prevent violating the SLO to a small degree for 5\% of the requests, it reacts promptly to the spike and is able to stabilize the latency after a few seconds to a similar level as before the spike.
Although \coolName{} does violate the SLO for the 5\% of requests in the tail, its prompt reaction to the spike allows it to quickly bring down the latency to the pre-spike levels. The longest SLO violation observed with \coolName{} is 35 seconds (Online Boutique, Trace B). For others, the violations last from 9 to 15 seconds.
Without \coolName, SLO violations last for well over a minute for all three applications: 1.45-2.3 minutes for Trace A and 1.4-2 minutes for Trace B. 
\cref{table:latency_reduction} summarizes all SLO violations for the VM-only baseline and \coolName.

In summary, \coolName~ is highly effective in absorbing sudden load spikes. It eliminates all SLO violations for the median case and exceeds the SLO at the tail for at most 35s, with actual latencies lower by 32-60.7\% (average 49.7\%) as compared to the VM-only baseline.

\begin{table}[t]
% 1. Increase row height (1.0 is default, 1.5 is 50% larger)
\renewcommand{\arraystretch}{1.3} 
\footnotesize
% \setlength{\belowcaptionskip}{8pt}
% 2. Optional: Increase column padding (6pt is default)
% \setlength{\tabcolsep}{8pt}
\centering
\begin{tabular}{|l|l|c|c|c|c|}
\hline
\multirow{2}{*}{\textbf{Application}} & \multirow{2}{*}{\textbf{Trace}} &  \multicolumn{2}{c|}{\textbf{Median (P50)}} & \multicolumn{2}{c|}{\textbf{Tail (P95)}} \\ \cline{3-6}
                                      &       & \textbf{VM} & \textbf{\coolName{}} & \textbf{BL} & \textbf{\coolName{}} \\ \hline
\multirow{2}{*}{\textbf{BookInfo}}    & \textbf{A}       & 36s   & --  & 1.45m & 12s    \\ \cline{2-6}
                                      & \textbf{B}       & 30s   & --  & 1.4m &  9s    \\ \hline
\multirow{2}{*}{\textbf{Online Boutique}} & \textbf{A}   & 3s    & --  & 1.6m &  10s  \\ \cline{2-6}
                                          & \textbf{B}   & 59s   & --  & 2m &  35s \\\hline
\multirow{2}{*}{\textbf{Hotel Reservation}} & \textbf{A} & 29s   & --  & 2.3m &  13s \\ \cline{2-6}
                                            & \textbf{B} & 25s   & --  & 1.6m &  15s \\ \hline
\end{tabular}
\caption{Service-Level-Agreement (SLO) violations. The time (in seconds) the end-to-end latency exceeds 400ms for Baseline (BL) and \coolName{}}
\label{table:latency_reduction}
% \vspace{-1em}
\end{table}

% \begin{figure}[t]
%     \centering
%     \includegraphics[width=0.49\textwidth]{figures/coldwarm2.png}
%     \caption{Latency comparison with \coolName{} using AWS Lambda for serverless}
%     \label{fig:coldstarts}
% \end{figure}

\begin{figure}[!t]
\centering
\includegraphics[width=0.4\textwidth]{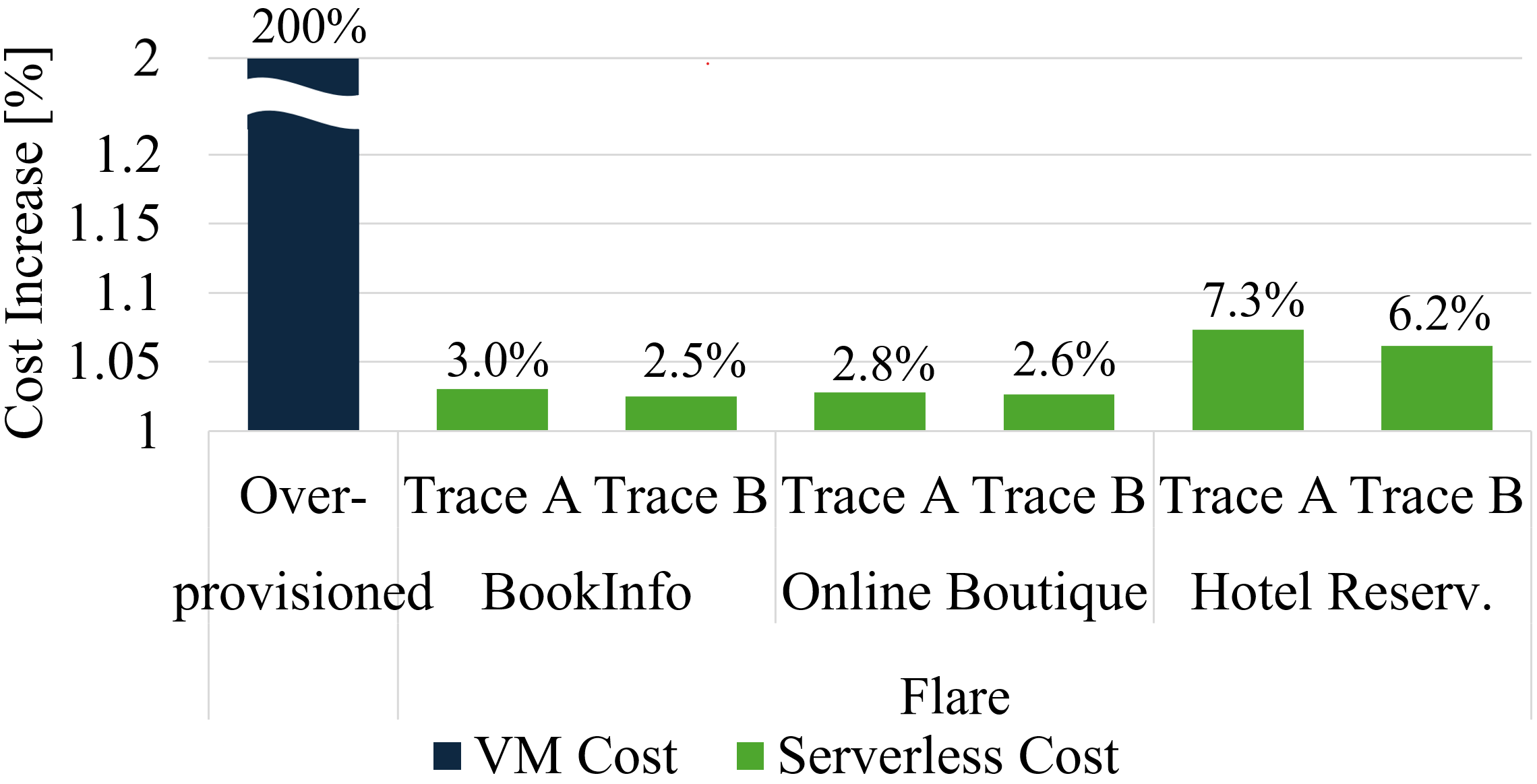}

\caption{Cost comparison}
\label{fig:eval-cost}
% \vspace{-1em}
\end{figure}

\subsection{Cost}
\label{sec:evaluation:cost}

Next, we evaluate the cost \coolName~ incurs over the VM-only baseline.  We compare \coolName{} against an over-provisioned VM-only cluster (\textit{over-provisioned}), with twice the resources of the baseline cluster, which we found is required to fully absorb the load spike in both traces.

% which can fully absorb the peak load. 
% . The over-provisioned cluster is configured to absorb the peak load in the studied trace 
% with an average CPU utilization of 50\% or less. As the observed peak load is approximately double the stable load, the over-provisioned configuration has twice the resources of the baseline cluster. 

\cref{fig:eval-cost} shows the results of the study, with results normalized to the baseline VM-only cluster.
% Throughout an hour-long trace, the baseline costs for the three distinct applications, namely BookInfo, Online Shop, and Hotel Reservation, are denoted as \dilina{\$XX, \$XX, and \$XX} correspondingly. In contrast, the costs incurred by \coolName~ demonstrate a \dilina{XX\%, XX, and XX}increase relative to the respective baseline costs across the three applications. Within these percentage increases, the serverless execution cost constitutes \dilina{XX,XX,XX}\%. 
For Trace A, the relative cost for utilizing \coolName~ compared to the baseline increases by 3\%, 2.8\%, and 7.3\% (4.4\% on average) across the three evaluated applications (BookInfo, Online Shop, and Hotel Reservation, respectively). Similarly, Trace B's increases stand at 2.5\%, 2.6\%, and 6.1\% (3.8\% on average) for the same applications. This indicates that \coolName~ can effectively absorb load spikes with very modest cost impact (less than 4.1\% on average).

\subsection{Evaluation on AWS Lambda}
\label{sec:evaluation:lambda}

% We study the effect of the load spikes described in \cref{sec:methodology} on the BookInfo microservice application, 
% \david{\soutWe demonstrate \coolName’s effectiveness using a popular production-ready serverless offering, AWS Lambda, under two scenarios: keeping serverless instances warm and cold. We observe the impact of load spikes described in \cref{sec:methodology} on tail latency using BookInfo microservice application.}}
We evaluate \coolName's effectiveness in handling load spikes described in \cref{sec:methodology} using a popular production-ready serverless offering, AWS Lambda, under two scenarios: (1) keeping serverless instances warm and (2) with cold instances that are provisioned on-demand by the serverless environment (Lambda). We perform the study using the BookInfo application.

%\textbf{Cold starts.} In serverless computing, the term cold starts refers to the delay experienced when a serverless function is invoked for the first time or after a period of inactivity. Unlike traditional server-based models, where servers are continuously running and ready to process requests, serverless platforms dynamically allocate resources only when a function is triggered. During a cold start, the serverless platform must initialize the necessary runtime environment, load the function code, and potentially allocate resources, which can result in increased latency for the first invocation. Despite efforts to reduce cold start delays, they remain a significant concern as they can introduce latency and affect user experience \cite{manner:cold-start,vahidinia:cold-start,baldini:serverless-computing}. Thus, we evaluate the effectiveness of \coolName~ under cold starts. 

\begin{figure}[t]
    \centering
    \includegraphics[width=0.49\textwidth]{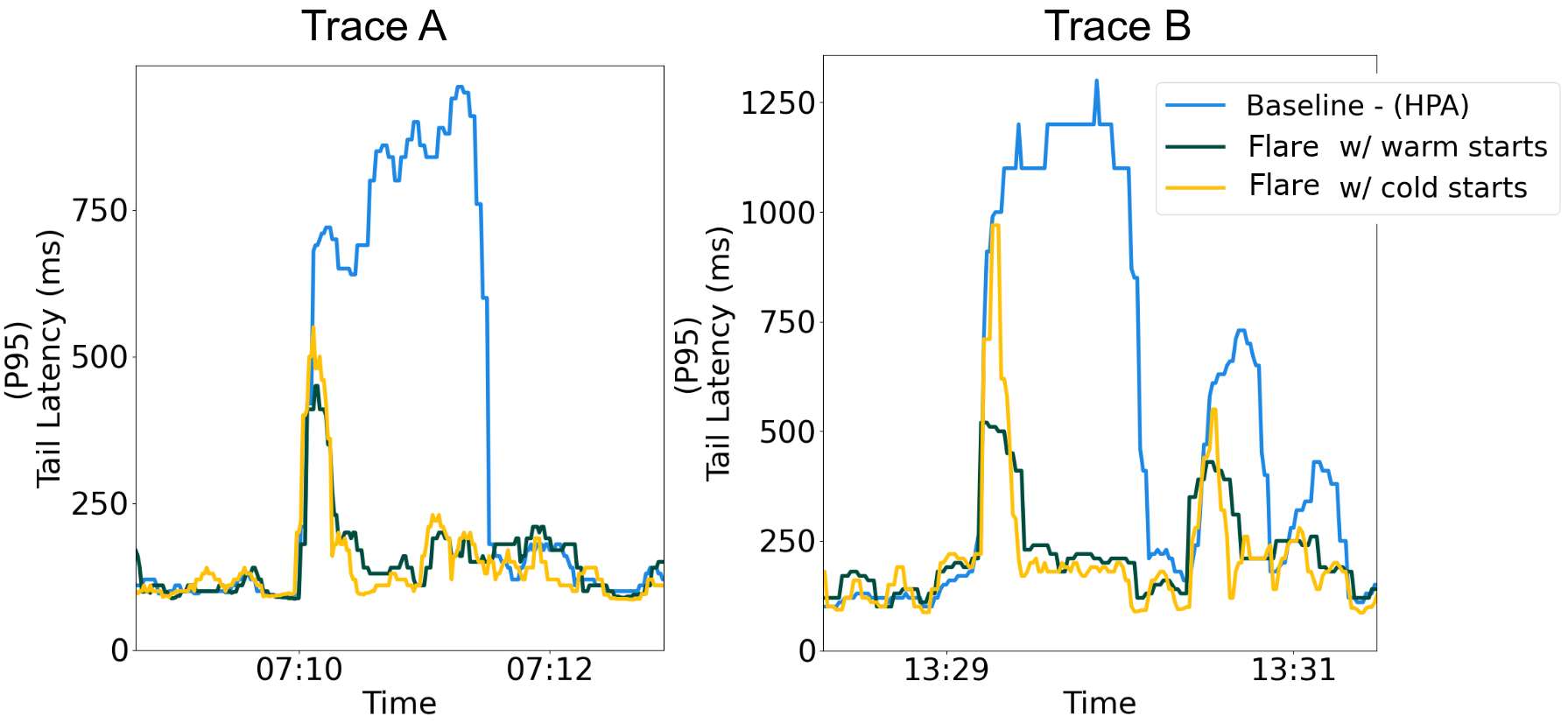}
    \caption{Latency comparison with \coolName{} using AWS Lambda for serverless}
    \label{fig:coldstarts}
\end{figure}

\cref{fig:coldstarts} illustrates the result of the study. The performance of \coolName~ with warm instances of AWS Lambda is similar to the results from evaluation shown in \cref{sec:evaluation:perf}, which also relied on warm instances albeit on Knative.
% Without \coolName{} tail latency spikes by more than 8x for Trace A and 12x for Trace B for ??s and ??s, respectively. 
By keeping serverless instances warm, \coolName{} is able to reduce the peak tail latency by 53\% and 61\% for Trace A and Trace B, respectively. Furthermore, the spike is quickly absorbed after 9s and 21s for Traces A and B, respectively.

With cold instances, \coolName~ exhibits 1.19x and 1.86x tail latency increases compared to \coolName~ with warm function instances for traces A \& B, respectively. The higher latencies are expected since instances must be provisioned on-demand when the spike traffic arrives. Notably, however, latencies with cold instances return to pre-spike levels just as fast as they do with warm instances. For \coolName~with both warm and cold instances, latencies return to pre-spike levels within a handful of seconds (10-16s), as compared to over a minute (1.45 minutes for Trace A and 1.4 minutes for Trace B) for the VM-only baseline.

%Compared to pre-spike latencies, \coolName~ with warm instances, cold instances, and the baseline show a latency increase of 4.2x, 5.2x, and 9.1x for trace A, and 4.9x, 9.2x, and 12.3x for trace B, respectively. In conclusion, this study demonstrates that \coolName~ remains effective in mitigating tail latency even in today's production serverless deployments, including cold start scenarios.

\begin{figure}[t]
    \centering
    \includegraphics[width=0.49\textwidth]{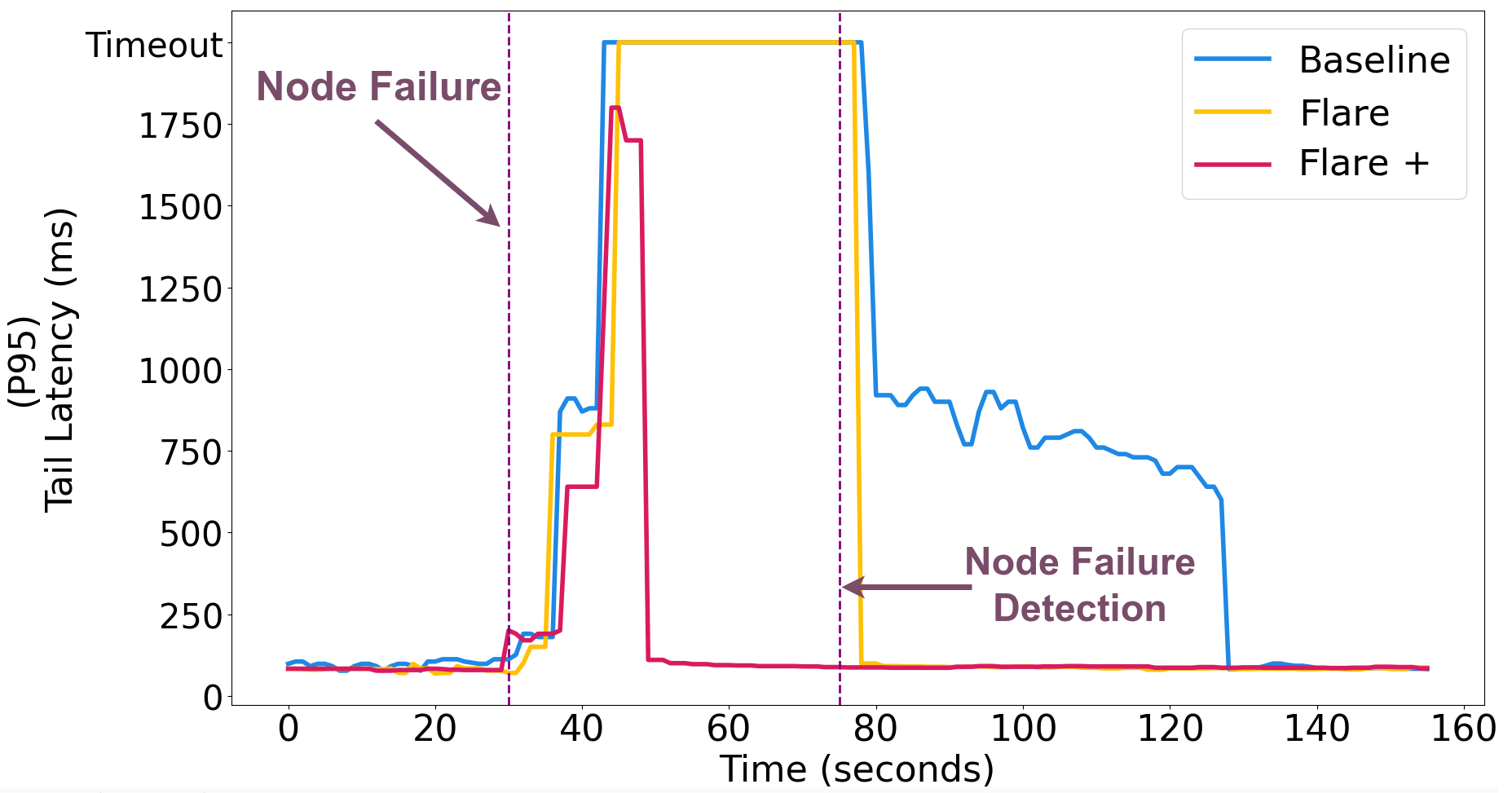}
    \caption{Impact of a node failure on tail latency}
    \label{fig:bookinfo-eval-nodefail}
\end{figure}

\subsection{\coolName{} for Node Failure Resilience}
\label{sec:evaluation:failures}
While \coolName{}'s primary objective is absorbing unexpected load spikes, it also offers an effective mechanism for mitigating node failures, which are a common occurrence in clusters due to hardware, network, or software faults~\cite{google-distributed}. In a traditional architecture, losing a node causes severe service degradation because K8s relies on a conservative default failure detection mechanism that takes 45 seconds to trigger~\cite{k8s-node-heartbeat}, and provisioning a replacement VM takes significantly longer. 

We evaluate this use-case of \coolName{}~on the BookInfo application deployed in an EKS cluster and serving a constant load.
In the experiment, the first service (called ProductPage) in the microservice chain is deployed on four VM-based instances. 30 seconds into the experiment, two of the VMs are terminated, mimicking, for instance, a network fault. As illustrated in \cref{fig:bookinfo-eval-nodefail}, when two nodes are artificially terminated, the baseline VM-only deployment suffers continuous timeouts during this detection window. Once detected, the load redistributes to the surviving nodes, overwhelming them and causing a 10x spike in tail latency that lasts for an additional 55 seconds. \coolName{}, however, immediately redirects the displaced traffic to serverless instances, stabilizing tail latencies within 5 seconds of failure detection. Furthermore, if the K8s detection threshold is reduced to 1 second (\textit{\coolName{}+}), tail latency returns to pre-spike levels just 18 seconds after the fault occurs. This demonstrates that selectively offloading traffic to serverless functions provides a robust safety net for infrastructure faults without the need for continuous over-provisioning.

\section{Related work}

\beginbsec{Reactive auto-scaling}
Reactive auto-scaling is well studied in the context of cloud-computing \cite{autoscale, reactive1,autoscaling-study}. Auto-scaling in public-clouds is primarily rule based where the rules are pre-defined by the cloud tenant, usually combining a resource with its utilization  \cite{awsScaling, azureScaling, googleScaling}. Rule-based auto-scaling struggles to timely react to large and sudden changes in load due to the delays in detecting the change and in reacting to it by launching more VMs. While ATOM \cite{atom} and Microscaler \cite{microscaler} improve scaling accuracy using queueing models, they remain bound by underlying VM initialization latency.

\beginbsec{Proactive auto-scaling}
Proactive auto-scaling techniques aim to scale resources by predicting future load changes, thereby eliminating the detection delay that hinders reactive auto-scaling schemes. Prior works~\cite{roy2011efficient, luo2022power, ali2023hybrid,predictive-rl} leverage past data to anticipate variations in load patterns. However, these approaches rely heavily on historical patterns or offline training, rendering them ineffective against unprecedented or non-periodic spikes (anomalies) which have not been encountered before. Furthermore, predictive models gracefully complement \coolName: the predicted load can guide the baseline VM pool size, while \coolName~ serves as a safety net to handle unexpected load spikes that exceed the predicted capacity.

\beginbsec{Fast-booting infrastructure}
Emerging virtualization technologies like MicroVMs \cite{firecracker}, Unikernels \cite{unikernal-cloud,unikernals-2}, and snapshot-based restoration \cite{snapshot-restore,restore-microvm} are designed to minimize boot times to the order of milliseconds, addressing the cold start latency often associated with standard heavy-weight VMs. For instance, the most popular cloud provider, AWS, runs its serverless offering (Lambda) on Firecracker MicroVMs \cite{firecracker} to enable rapid elasticity. \coolName{} is designed to be agnostic to this underlying compute, requiring only that the secondary tier provides fast provisioning. While our evaluation utilizes serverless functions due to their widespread availability, \coolName{}'s control plane can seamlessly support other fast-booting compute types as they become standard cloud offerings.

\beginbsec{Combining multiple cloud service tiers}
Prior work has used Burstable VM instances \cite{burstable-instances-aws, burstable-instances-azure} to handle unexpected load peaks \cite{burscale,autoburst}. Burstable instances are VMs that provide a baseline performance (like a normal VM) plus burst performance for a short duration when needed. Burstable instances operate by keeping some idle CPU capacity on reserve inside a provisioned VM. While this makes burstable instances quick to engage in case of a sudden load spike, they are inherently limited by the degree of scaling they can sustain. Cloud providers also limit the duration for which these instances can burst. This combination limits the effectiveness of burstable instances in the face of large and sudden load spikes.
In contrast, \coolName~ uses serverless functions to handle the load spikes. Unlike burst instances, serverless functions do not need to be provisioned beforehand and have nearly unlimited scaling capacity. Moreover, \coolName{} can be used together with burst instances to provide a safety net in case the burst instance capacity is insufficient to accommodate a load spike. 

% \beginbsec{Hybrid systems with serverless}
% Prior works have used serverless in various contexts to handle load spikes \cite{splitserve, mark, cackle, pixel, libra, spock, feat}. 
% SplitServe \cite{splitserve} and Cackle \cite{cackle} use serverless functions to accommodate load spikes in data analytics workloads. However, they profile the workload ahead of time and leverage the profile to offload complete call-graphs at query arrival time. In a microservice architecture, which is the focus of this work, the call graphs are request-dependent and can even change for different input values of the same request type. \coolName~ is designed to handle such dynamic call graphs. 
% Pixel~\cite{pixel} also focuses on analytical workloads but offloads only one type of operator (Filter) to serverless, relying on a-priori query knowledge. 
% Similar to analytical workloads, Feat \cite{feat} only focuses on fixed call-graphs of publish-subscribe model.
% Mark, Libra and Spock~\cite{mark, libra, spock} leverage different cloud offerings including serverless for ML inference-as-a-service. Each worker instance in these works is a stand-alone inference engine.
% Unlike \coolName{}, none of these works focus on latency-sensitive applications that comprise of a dynamic chain of microservices.

\beginbsec{Hybrid systems with serverless}
Prior works have used serverless in various contexts to handle load spikes~\cite{splitserve, cackle, sponge, mashup, mark, spock, libra, feat, beehive}. 
Systems such as SplitServe~\cite{splitserve}, Cackle~\cite{cackle}, Sponge~\cite{sponge}, and Mashup~\cite{mashup} utilize serverless functions to accommodate load spikes in data analytics, stream processing, and high-performance computing workloads. These systems target execution models where query, stream, or workflow structure is exposed to the runtime, and they rely on workload profiling, compile-time graph rewriting, or cost models to decide how work should be split across VMs and serverless resources.

Other works explore hybrid provisioning for machine learning inference-as-a-service~\cite{mark, spock} and general cloud applications with strict SLAs~\cite{libra, feat, beehive}. A fundamental characteristic of these systems is that they treat workloads as isolated tasks or monolithic units. For instance, serverless instances in Mark~\cite{mark} and Spock~\cite{spock} operate as stand-alone inference engines evaluating independent requests. Similarly, Libra~\cite{libra} and Feat~\cite{feat} use serverless instances primarily to handle bursty frontend traffic or as interim resources to compensate for VM launch delays, operating at the application's entry point for traffic. BeeHive~\cite{beehive} offloads parts of a monolithic web application to FaaS. These systems are distinct from \coolName{}, which shifts traffic for specific bottlenecked services in a multi-hop microservice chain. Furthermore, a key strength of \coolName{} is its practical deployability; it offers easy, non-intrusive compatibility with the industry-grade service meshes already prevalent in current microservice deployments.

\section{Conclusion}
Dealing with sudden spikes in load poses a considerable challenge for existing VM-based microservices due to the time required to bring up additional resources. Traditional VM-based reactive auto-scaling struggles to maintain strict SLOs during sudden load spikes due to inherent detection and reaction lags while predictive scaling cannot reliably foresee unexpected surges. Although serverless functions with ultra-fast startup times offer a promising solution, the high cost of a serverless-only approach hinders widespread adoption. 
To address this issue, we propose \coolName{}, a hybrid architecture that seamlessly combines VM based microservices with serverless computing to improve scalability and load resilience. \coolName{} dynamically identifies overloaded services and selectively routes only the excess traffic to serverless functions. Our evaluation shows that \coolName{} is highly effective in absorbing sudden load spikes. It eliminates all SLO violations for the median case and only slightly exceeds the SLO for the tail at a marginal cost increase of 4.1\%, on average.  

% \printbibliography
\bibliographystyle{IEEEtranN}
\bibliography{bib}
\end{document}